\def\gtrsim{\begin{array}{c} > \\ \sim \end{array}}
\def\lesssim{\begin{array}{c} < \\ \sim \end{array}}

\documentclass[
    ,final            
  ]
  {aipproc}

\layoutstyle{6x9}

\begin{document}

\title{Spacetime Foam and Dark Energy}

\classification{04.60.Bc, 95.36.+x}

\keywords      {spacetime foam, quantum foam, dark energy, holography, infinite 
statistics, nonlocality}

\author{Y. Jack Ng}{
  address={Institute of Field Physics, Department
of Physics and Astronomy, University of North Carolina, Chapel
Hill, NC 27599, USA}
}

\begin{abstract}

Due to quantum fluctuations, spacetime is foamy on small scales.  The degree
of foaminess is found to be consistent with the holographic principle.
One way to detect
spacetime foam is to look for halos in the images of distant quasars.
Applying
the holographic foam model to cosmology we
"predict" that the cosmic energy density takes on the critical value; and 
basing only on existing archived data on active galactic
nuclei from the Hubble Space Telescope, we also
"predict" the existence of dark energy which, we argue, is composed of an
enormous number of inert ``particles" of extremely long wavelength.  We
speculate that these ``particles" obey infinite statistics.



\end{abstract}

\maketitle


\section{Introduction and Summary}

Like everything else, spacetime is conceivably subject to quantum
fluctuations.  So we expect that spacetime, probed at a small enough scale,
will appear complicated --- something akin in complexity to a turbulent
froth that John Wheeler
has dubbed "quantum foam," also known as "spacetime
foam."\footnote{In the gravitational context, the phenomenon of turbulence is
indeed intimately related to the properties of spacetime foam.  See Ref.
\cite{Jejjala2008}.}  But how large are the fluctuations in the fabric of
spacetime?
To quantify the problem, let us recall that, if spacetime indeed
undergoes quantum fluctuations, there will be an intrinsic limitation to
the accuracy with which one can measure a distance, for that distance
fluctuates.  Denoting the fluctuation of a distance $l$ by $\delta l$, on
general grounds, we expect $\delta l \gtrsim l^{1 - \alpha} l_P^{\alpha}$,
where $l_P = \sqrt{\hbar G/c^3}$ is the Planck length,
the characteristic length scale in quantum gravity, and we have denoted
the Planck constant, gravitational constant and the speed of light by
$ \hbar$, $G$ and $c$ respectively.
The parameter $\alpha
\sim 1$ specifies the different spacetime foam models.  In this talk we will
concentrate on the model corresponding to $\alpha = 2/3$, 
which has come to be known as the holographic model \cite{wigner,Karol}, 
so called because it
is found to be consistent with the holographic principle \cite{tHooft,susskind},
according to which, the information content inside any three dimensional
region of space can be encoded on the two dimensional surface around the
region, like a hologram.
For comparison, we will also consider the random-walk model \cite{Amelino1999} 
corresponding to $\alpha = 1/2$.

Contents of this talk:
Applying nothing more than quantum mechanics and some rudimentary 
black hole physics,
we derive the holographic model of spacetime foam. 
Applying the holographic model to cosmology, we "predict" that the
cosmic energy density takes on the critical value (i.e., the 
fractional density parameter of the universe $\Omega \cong 1$),
consistent with observation.  Then
aided by some archived data on quasar or AGN from the Hubble Space
Telescope, we are led to conclude that dark energy exists.
Furthermore we are naturally led to
speculate that the constituents of dark energy, unlike ordinary matter,
obey an exotic statistics
known as infinite statistics in which all
representations of the particle permutation group can occur.

\section{Fluctuations of Spacetime and Models of Spacetime Foam}

One way to find out how much a distance $l$ fluctuates (i.e., the magnitude
of $\delta l$) is to carry out a gedanken experiment to measure
$l$.  But, for later use, it is
more convenient to find $\delta l$ by carrying out a process of mapping
the geometry of spacetime.  This method 
\cite{llo04}        
relies on the fact that quantum fluctuations of spacetime
manifest themselves in the form of uncertainties in the geometry of
spacetime.  Hence the structure of spacetime foam can be inferred from the
accuracy with which we can measure that geometry.  Let us
consider mapping out the geometry of spacetime for a spherical volume of
radius $l$ over the amount of time $T = 2l/c$ it takes light to cross the
volume.  One way to do this is to fill the space with clocks, exchanging
signals with other clocks and measuring the signals' times of arrival.
This process of mapping the geometry of spacetime is a kind of computation,
in which distances are gauged by transmitting and processing information.
The total number of operations, including the ticks of the clocks and
the measurements of signals, is bounded by the Margolus-Levitin theorem 
\cite{ML}
in quantum computation,
which stipulates that the rate of operations for any computer       
cannot exceed the amount of energy $E$ that is available for computation
divided by $\pi \hbar/2$.   A total mass $M$ of clocks then
yields, via the Margolus-Levitin theorem, the bound on the total number of
operations given by $(2 M c^2 / \pi \hbar) \times 2 l/c$.  But to prevent
black hole formation, $M$ must be less than $l c^2 /2 G$.  Together, these
two limits imply that the total number of operations that can occur in a
spatial volume of radius $l$ for a time period $2 l/c$ is no greater than  
$2 (l/l_P)^2 / \pi$.  To maximize spatial resolution, each clock must tick
only once during the entire time period.  The operations can be regarded as
partitioning the spacetime volume into "cells", then on the average each cell
occupies a spatial volume no less than $(4 \pi l^3 / 3) / (2 l^2 /\pi l_P^2)
=
2 \pi^2 l l_P^2 /3$, yielding an average separation between neighhoring   
cells
no less than $(2 \pi^2 /3)^{1/3} l^{1/3} l_P^{2/3}$.  This spatial      
separation is interpreted as the average minimum uncertainty             
in the measurement of
a distance $l$, that is, 
\begin{equation}
\delta l \gtrsim l^{1/3} l_P^{2/3}, 
\label{vD2}
\end{equation}
where and henceforth (with a couple of exceptions)                      
we drop multiplicative factors of order $1$.
(Recently Gambini and Pullin 
\cite{Gambini2007} 
have derived from first principles, in
the framework of loop quantum gravity in spherical symmetry,
an uncertainty in the
determination of volumes that grows radially, 
consistent with Eq. \eqref{vD2}.)

We can now understand why this quantum foam model has come 
to be known
as the holographic model.
Since, on the average, each cell occupies a spatial volume of $l l_P^2$,  
a spatial region of size $l$ can contain no more than $l^3/(l l_P^2) =  
(l/l_P)^2$ cells.  Thus this model
corresponds to the case of
maximum number of bits of information $l^2 /l_P^2$
in a spatial region of size $l$, that is
allowed by the holographic principle. 

It will prove to be useful to compare the holographic model in the mapping
of the geometry of spacetime
with the one that corresponds to spreading the spacetime cells uniformly
in both space and time.  For the latter case, each cell has
the size of $(l^2 l_P^2)^{1/4} =
l^{1/2} l_P^{1/2}$ both spatially and temporally so that each clock ticks
once in the time it takes to communicate with a neighboring clock.  Since
the dependence on $l^{1/2}$ is the hallmark of a random-walk fluctuation,
this quantum foam model corresponding to  $\delta l \gtrsim  
(l l_P)^{1/2}$ is called the random-walk model. 
Compared to the holographic model, the random-walk model predicts a     
coarser spatial resolution, i.e., a larger distance fluctuation.  It       
also yields a smaller bound on the information content in a spatial
region, viz., $(l/l_p)^2 / (l/l_P)^{1/2} = (l^2 / l_P^2)^{3/4} =       
(l/l_P)^{3/2}$ bits.

Note that the minimum $\delta l$ just found for the
holographic model corresponds
to the case of maximum energy density $\rho = (3/ 8 \pi) (l l_P)^{-2}$    
(in units $c=1=\hbar$, and Boltzmann constant $k_B=1$ for later use)
for a sphere of radius $l$ not
to collapse into a black hole.  Hence the holographic model, unlike     
the other models, requires, for
consistency, the energy density to have the critical value.
(By contrast, for instance, one can show that 
the  random-walk model corresponds to an energy
density
$(l l_P)^{-2}\gtrsim \rho \gtrsim l^{-5/2} l_P^{-3/2}$.)

\section{Probing Spacetime Foam and the Fall of the Random-Walk Model}

The Planck length $l_P \sim 10^{-33}$ cm is so short that we need an
astronomical (even cosmological) distance $l$ for its fluctuation $\delta
l$ to be detectable.
Let us consider light (with wavelength $\lambda$)             
from distant quasars or bright active galactic nuclei. 
Due to quantum fluctuations of spacetime, the wavefront, while planar,
is itself ``foamy", having random fluctuations in phase 
$\Delta \phi \sim 2 \pi \delta l / \lambda$ and in       
the direction of the wave vector 
given by $\Delta \phi / 2 \pi
$ (for $\delta l \ll \lambda$).  In effect, spacetime foam creates a
``seeing disk" whose angular diameter is $\sim \Delta \phi /2 \pi  $.  For
an interferometer with baseline length $D$, this means that dispersion will
be seen as a spread in              
the angular size of a distant point source, causing a reduction in the
fringe
visibility when $\Delta \phi / 2 \pi \sim \lambda / D  $. For a quasar of 1
Gpc away, at infrared wavelength, the
holographic model predicts a phase fluctuation $\Delta \phi \sim 2 \pi
\times      
10^{-9}  $ radians.  On the other hand, an infrared interferometer (like the
Very Large Telescope Interferometer) with $D \sim 100$ meters has $\lambda /
D \sim 5 \times 10^{-9}  $.
Thus, in principle, this method will allow the use of interferometry fringe
patterns to test the holographic model!  (For more  
discussion of this proposal to detect spacetime foam, see Ref. 
\cite{Christiansen2008}.)

In the meantime, we can use existing archived data on                 
quasars or active galactic nuclei
from the Hubble Space Telescope to test the quantum foam models.
Consider, for example, the case
of PKS1413+135, 
an AGN for which the redshift is $z = 0.2467$. \cite{per02}
With $l \approx 1.2$ Gpc and $\lambda = 1.6 \mu$m, we 
find $\Delta \phi \sim 10 \times 2 \pi$ and
$10^{-9} \times 2 \pi$ for the random-walk model and
the holographic model of spacetime foam respectively.
With $D = 2.4$ m for HST, we expect to detect halos
if $\Delta \phi \sim 10^{-6} \times 2 \pi$.
Thus, the HST image only fails to test the holographic model by
3 orders of magnitude. 

However, the absence of a quantum foam induced halo structure in the      
HST image of PKS1413+135 rules out convincingly
the random-walk model.  (In fact, the scaling relation  
discussed above indicates             
that all spacetime foam models with $\alpha \lesssim 0.6$ are ruled
out by this HST observation.)

\section{From Spacetime Foam to Cosmology and "Prediction" of Dark Energy}

Assuming that there is a unity of physics connecting the Planck scale to the
cosmic scale, we can now appply the holographic spacetime foam model to 
cosmology \cite{llo04,Arzano,plb} and henceforth we call that cosmology 
the holographic foam cosmology (HFC).  The fact that our
universe is observed to be at or very close to its critical density
must be taken as solid albeit indirect evidence in favor of the
holographic model because, as discussed above, it is the only      
model that requires, for its consistency,  
the maximum energy density without causing gravitational collapse.
Specifically, according               
to the HFC, the cosmic density is
\begin{equation}                           
\rho = (3 / 8 \pi) (R_H l_P)^{-2} \sim (H/l_P)^2,
\label{density}
\end{equation}
where $H$ is the Hubble parameter of the observable universe
and $R_H$ is the Hubble radius,
and the cosmic entropy is given by 
\begin{equation}
I \sim (R_H /l_P)^2.
\label{bits}
\end{equation}

For the present cosmic era, the energy density is given by
$\rho \sim H_0^2/G \sim (R_H l_P)^{-2}$ (about $10^{-9}$  
joule per cubic meter).
Treating the whole universe as a computer,
one can
apply the Margolus-Levitin theorem to conclude that the universe
computes at a rate $\nu$ up to $\rho R_H^3 \sim R_H l_P^{-2}$
($\sim 10^{106}$ op/sec), for a total of $(R_H/l_P)^2$              
($\sim10^{122}$) operations during its lifetime so far.
If all the information of this huge computer is stored in ordinary
matter, we can apply standard methods of statistical mechanics
to find that the total number $I$ of bits is $[(R_H/l_P)^2]^{3/4} =
(R_H/l_P)^{3/2}$ ($\sim 10^{92}$).
It follows that each bit flips once in the amount of time given by
$I/\nu \sim (R_H l_P)^{1/2}$ ($\sim 10^{-14}$ sec).                    
However the average separation of neighboring bits is
$(R_H^3/I)^{1/3} \sim (R_H l_P)^{1/2}$ ($\sim 10^{-3}$
cm).  Hence, assuming only ordinary matter exists to store all the
information we are led to conclude that the time
to communicate with neighboring bits is equal to the time for each
bit to flip once.  It follows that the accuracy to which ordinary
matter maps out the geometry of spacetime corresponds exactly to            
the case of events
spread out uniformly in space and time discussed above for the case
of the random-walk model of spacetime foam.       

But, as shown in the previous section,
the sharp images of distant quasars or active galactic nuclei
observed at the Hubble Space
Telescope have ruled out the random-walk model. From the demise of the
random-walk model and the fact that ordinary matter only contains an 
amount of information dense enough to map out spacetime at a level
consistent with the random-walk model, one now infers that
spacetime is mapped to a finer spatial accuracy than that which
is possible with the use of ordinary matter.  
Therefore there must be another kind of substance with which spacetime
can be mapped to the observed accuracy, conceivably as given       
by the holographic model. The natural conclusion
is that unconventional
(dark) energy/matter exists!
Note that this argument does not make use of                             
the evidence from recent cosmological (supernovae, cosmic microwave
background, and galaxy clusters) observations. 

Furthermore, from Eqs. \eqref{density} and \eqref{bits},
the average energy carried by each particle/bit of the unconventional
energy/matter is
$\rho R_H^3/I \sim R_H^{-1}$ ($\sim 10^{-31}$ eV).  Such
long-wavelength (hence ``non-local'')
bits or ``particles'' carry negligible kinetic energy.
Also according to HFC, it takes each unconventional bit
the amount of time
$I/\nu \sim R_H$ to flip.  Thus, on the average, each bit flips
once over the course of cosmic history.  Compared to the conventional       
bits carried by ordinary matter, these bits are rather passive and      
inert (which, by the way, may explain why dark energy is dark).

\section{Dark Energy, Holography, Infinite Statistics and Nonlocality}

What is the overriding difference between conventional matter and
unconventional energy/matter (i.e., dark energy and perhaps also dark
matter)?
To find that out, let us 
consider a perfect gas of $N$ particles obeying Boltzmann statistics
at temperature $T$ in a volume $V$.  For the
problem at hand, let us take $V \sim R_H^3$, $T \sim R_H^{-1}$, and $N \sim
(R_H/ l_P)^2$. A standard calculation (for the relativistic case) yields the
partition function $Z_N = (N!)^{-1} (V / \lambda^3)^N$, where           
$\lambda = (\pi)^{2/3} /T$.
With the free energy given by
$F = -T ln Z_N = -N T [ ln (V/ N \lambda^3) + 1]$,
we get, for the entropy of the system,               
$S = - ( \partial F / \partial T)_{V,N} = N [ln (V / N \lambda^3) + 5/2]$.
The important point to note is that, since $V \sim \lambda^3$, the entropy
$S$ becomes nonsensically negative unless $ N \sim
1$ which is equally nonsensical because $N$ should not be too different from
$(R_H/l_P)^2 \gg 1$.  But the solution 
is pretty obvious: the $N$ inside the log in $S$ somehow
must be absent.  Then $ S \sim N
\sim (R_H/l_P)^2$ without $N$ being small (of order 1) and S is non-negative
as physically required.  That is the case if the ``particles" are
distinguishable and nonidentical!  For in that case, the Gibbs $1/N!$ factor
is absent from the partition function $Z_N$, and the entropy becomes
\begin{equation}
S = N[ln (V/ \lambda^3) + 3/2].
\label{entropy2}                              
\end{equation}

Now the only known consistent statistics in greater than two space
dimensions
without the Gibbs factor 
is infinite statistics (sometimes called
``quantum Boltzmann statistics"). 
\cite{Haag,greenberg}.  
Thus we have
shown that the ``particles" constituting dark energy obey infinite          
statistics,                    
instead of the familiar Fermi or Bose statistics. 
\cite{plb}.          
(Using the Matrix theory approach,
Jejjala, Kavic and Minic 
\cite{minic2007} 
have also argued that dark energy       
quanta obey infinite statistics.)
What is infinite statistics?  Succinctly, a Fock realization of infinite
statistics is provided by a $q$ deformation of the commutation relations of
the oscillators:
$a_k a^{\dagger}_l - q a^{\dagger}_l a_k = \delta_{kl}$ with $q$ between -1
and 1 (the case $q = \pm 1$ corresponds to bosons or fermions).  Two states      
obtained by acting with the $N$ oscillators in different orders are
orthogonal; i.e., the states may be in any representation        
of the permutation group.  

Infinite statistics appears to have one ``defect": a theory of particles
obeying infinite statistics cannot be local. 
\cite{greenberg}.
(That is, the fields associated with infinite statistics are not local,
neither in the sense that their observables commute at spacelike separation 
nor in the sense that their observables are pointlike functionals of the
fields.) 
Remarkably, the TCP theorem and cluster   
decomposition have been shown to hold despite the lack of locality.          
\cite{greenberg}.
Actually this lack of locality may have a
silver lining.  According to the holographic principle, the
number of degrees of freedom in a region of space is bounded not by
the volume but by the surrounding surface.  This suggests that the
physical degrees of freedom are not independent but, considered         
at the Planck scale, they must be infinitely correlated, with the result
that the spacetime location of an event may lose its invariant significance.
Since the holographic principle is believed to be
an important ingredient in the formulation of quantum gravity,              
the lack of locality for theories of infinite statistics may not be a   
defect; it can actually be a virtue. 
Quantum gravity and infinite statistics appear to fit together
nicely, and the nonlocality present in systems obeying infinite statistics 
may be related to the nonlocality present in holographic theories. 


But there is the question whether cosmic energy 
density $\rho \sim H^2/G$ can lead to the accelerating cosmic 
expansion as observed.  Fortunately, it has been pointed out                
by Zimdahl and Pavon 
\cite{Zimdahl2006}
that a transition from
an earlier decelerating to a recent and present 
accelerating cosmic expansion can arise as a pure interaction       
phenomenon if dark matter is coupled to holographic
dark energy with $\rho \propto H^2$. 
As a bonus, within the framework of such cosmological models,               
we can now understand
why, in addition to dark energy, dark matter has to exist.  
However the phenomenology of holographic
foam cosmology has yet to be worked out in detail; further work 
in this area is warranted.

\begin{theacknowledgments}

This work was supported in part by the US Department of Energy. 
I thank L.~L. Ng for his help in the preparation of the manuscript.
This talk is largely based on a recent review article \cite{Ng2008}
which contains a considerably more complete set of references.  

\end{theacknowledgments}

\end{document}